# Different types of conditional expectation and the Lüders - von Neumann quantum measurement

Gerd Niestegge
Zillertalstrasse 39, D-81373 Muenchen, Germany

Abstract. In operator algebra theory, a conditional expectation is usually assumed to be a projection map onto a sub-algebra. In the paper, a further type of conditional expectation and an extension of the Lüders - von Neumann measurement to observables with continuous spectra are considered; both are defined for a single operator and become a projection map only if they exist for all operators. Criteria for the existence of the different types of conditional expectation and of the extension of the Lüders - von Neumann measurement are presented, and the question whether they coincide is studied. All this is done in the general framework of Jordan operator algebras. The examples considered include the type I and type II operator algebras, the standard Hilbert space model of quantum mechanics, and a no-go result concerning the conditional expectation of observables that satisfy the canonical commutator relation.

Key words: conditional expectations, quantum measurement, operator algebras, Jordan algebras

## 1. Introduction

In operator algebra theory, a conditional expectation is usually assumed to be an idempotent positive linear map onto a sub-algebra[2,4,10,11,12]. The applicability of such conditional expectations to the Lüders - von Neumann measurement in quantum mechanics has been studied by Areveson[1]. The intention was to extend the Lüders - von Neumann measurement to observables with continuous spectra, but the result is negative - at least for the quantum mechanical standard model using the algebra of all bounded linear operators on a separable Hilbert space. In other cases (e.g., the type $II_1$ factors), such an extension is possible.

A new type of conditional expectation that need not be a map on the whole operator algebra, but can be defined for each single operator has recently been introduced in Ref. [8]. The present paper studies both types of conditional expectation as well as an extension of the Lüders - von Neumann measurement to observables with continuous spectra. Criteria for their existence are presented and the question when they coincide is investigated. As an example, the canonical commutator relation $[X,Y]=i$ in the standard model of quantum mechanics is considered and it is shown that a conditional expectation of $Y$ under $X$ does not exist for such an observable pair.

The new type of conditional expectation is based on the concept of quantum conditional probabilities presented in Ref. [5]. Since an abelian Jordan product ∘ (but not the usual operator product) can be reconstructed from these quantum conditional probabilities, Jordan operator algebras and particularly the so-called JBW algebras [which are the Jordan analogue of the $W^*$-algebras or von Neumann algebras] are the appropriate framework for this approach[6]. The idempotent elements $E$ of a JBW algebra $\mathcal{A}$ then form the events, the negation of an event $E$ is its orthogonal complement $E'=\mathbb{1}-E$, where $\mathbb{1}$ is the unit of $\mathcal{A}$, and the states are the positive linear functionals $\mu$ with $\mu(\mathbb{1})=1$.

With a given state $\mu$, a conditional probability under an event $F$ with $\mu(F)>0$ is defined as a state $\nu$ with $\nu(E)=\mu(E)/\mu(F)$ for all those events $E$ with $E\leq F$, and we then write $\mu(Y|F):=\nu(Y)$ for $Y \in \mathcal{A}$. In a JBW algebra, the conditional probability $\mu(Y|F)$ becomes identical with $\mu(\{E,Y,E\})/\mu(F)$, where $\{X,Y,Z\}$ is the so-called Jordan triple product defined as $\{X,Y,Z\}=X \circ (Y \circ Z)-Y \circ (Z \circ X)+Z \circ (X \circ Y)$ for $X,Y,Z \in \mathcal{A}$. When $\mathcal{A}$ is the self-adjoint part of a $W^*$-algebra or von Neumann algebra, the Jordan product is $X \circ Y=(XY+YX)/2$; then $\{X,Y,Z\}=(XYZ+ZYX)/2$ and $\{X,Y,X\}=XYX$.

In the case when $\{E,F,E\}=\lambda E$ holds for two events $E$ and $F$ with some real number $\lambda$ (e.g., if $E$ is a minimal event), we have $\mu(F|E)=\lambda$ for all states $\mu$ with $\mu(E)>0$; this state-independent conditional probability is then denoted by $\mathbb{P}(F|E)$ and is a very special type of objective probability.[5]

The monograph by Hanche-Olsen and Størmer[2] is recommended as an excellent reference for the theory of Jordan operator algebras. The readers interested only in the von Neumann algebras or W*-algebras should keep in mind that $X \circ Y$ and $\{X,Y,X\}$ can be replaced by $(XY+YX)/2$ and $XYX$, respectively, in this case.

After an overview of the different types of conditional expectation in the next section, the extension of the Lüders - von Neumann measurement to observables with continuous spectra is introduced in Section 3. Section 4 then presents the main results dealing with the questions whether the different types of conditional expectation and the extended Lüders - von Neumann measurement exist and whether they are identical. The examples and applications considered in Section 5 include the type I and type II JBW factors, the extension of Areveson's no-go result to the type I JBW factors, and the canonical commutator relation in the quantum mechanical standard model.

## 2. Conditional expectations

Classical conditional probabilities always satisfy the identity $\mu(F)=\mu(F|E)\mu(E)+\mu(F|E')\mu(E')$ for any pair of events $E$ and $F$. However, this does not hold for the quantum conditional probabilities considered here and motivates the following definition[7]. With an event $E$ and any element $X$ in a JBW algebra $\mathcal{A}$ and a state $\mu$ on $\mathcal{A}$, we write $E \xrightarrow{\mu} X$ if the equation $\mu(X)=\mu(X|E)\mu(E)+\mu(X|E')\mu(E')$ or, equivalently, $\mu(\{E,X,E\}) = \mu(E \circ X)$ holds. With a JBW sub-algebra $\mathcal{B}$ of $\mathcal{A}$, we write $\mathcal{B} \xrightarrow{\mu} X$ if $E \xrightarrow{\mu} X$ holds for all events $E$ in $\mathcal{B}$, and we write $\mathcal{B} \xrightarrow{\mu} \mathcal{A}$, if if $E \xrightarrow{\mu} X$ holds for all events $E$ in $\mathcal{B}$ and all elements $X$ in $\mathcal{A}$. A state $\mu$ satisfying $\mathcal{A} \xrightarrow{\mu} \mathcal{A}$ is a finite trace.[9]

Lemma 2.1: Let $\mathcal{A}$ be a JBW algebra with unit $\mathbb{1}$, and let $\mathcal{B}$ be a JBW sub-algebra with $\mathbb{1} \in \mathcal{B}$. For any element $X \in \mathcal{A}$ and any state $\mu$ on $\mathcal{A}$, the following four conditions are equivalent:
(i) $\mathcal{B} \xrightarrow{\mu} X$
(ii) $\mu(\{E,X,E'\})=0$ for all events $E \in \mathcal{B}$.
(iii) $\mu(\{S,X,S\})=\mu(X)$ for all symmetries $S \in \mathcal{B}$ (A symmetry is an element $S$ with $S^2=S$).
(iv) $\mu(\{E,X,F\})=0$ for all orthogonal event pairs $E,F \in \mathcal{B}$.
If $\mathcal{A}$ and $\mathcal{B}$ are the self-adjoint parts of $W^*$-algebras, each of the above four conditions is equivalent to the following one:
(v) $\mu(UXU^{-1})=\mu(X)$ for all unitary elements $U$ in the $W^*$-algebra generated by $\mathcal{B}$.

Proof: The equivalence of (i) and (ii) follows from the identity $\{E,X,E'\}=E \circ X-\{E,X,E\}$. Each symmetry $S \in \mathcal{B}$ has the shape $E-E'$ with an event $E \in \mathcal{B}$. The identity $\{S,X,S\}=X-4\{E,X,E'\}$ then yields the equivalence of (ii) and (iii). Condition (iv) obviously implies (ii). We now prove (iv) assuming (i). Let $E,F \in \mathcal{B}$ be two orthogonal events. Then, $\mu(E \circ X)+\mu(F \circ X)=\mu((E+F) \circ X)$

- 2 -

$=\mu(\{E+F,X,E+F\})=\mu(\{E,X,E\})+\mu(\{F,X,F\})+2\mu(\{E,X,F\})=\mu(E\circ X)+\mu(F\circ X)+2\mu(\{E,X,F\})$, and therefore $\mu(\{E,X,F\})=0$.

We now assume that $\mathcal{A}$ is the self-adjoint part of a $W^*$-algebra. Condition (v) implies (iii), since symmetries are a special type of unitary elements with $S^{-1}=S=S^*$. We finally assume (i) and prove (v). Since any unitary element in the $W^*$-algebra generated by $\mathcal{B}$ can be approximated (in the norm topology on $\mathcal{A}$) by unitary elements with discrete spectrum, it is sufficient to consider $U=\Sigma\lambda_n E_n$ with $E_n$ being orthogonal events in $\mathcal{B}$ such that $\Sigma E_n = \mathbb{1}$ and with $\lambda_n$ being complex numbers such that $|\lambda_n|=1$. Then $\mu(UXU^{-1})=\Sigma_n\Sigma_m\lambda_n\lambda_m^{-1}\mu(E_n X E_m)=\Sigma_n\mu(E_n X E_n)=\Sigma_n\mu(E_n\circ X)=\mu(X)$, where first (iv) [which is implied by (i)] and then (i) itself have been used. □

For a normal state $\mu$, the condition $\mathcal{B}\xrightarrow{\mu}X$ is equivalent to the existence of a conditional expectation of $X$ under $\mathcal{B}$ in the state $\mu$; a conditional expectation is an element $Y\in\mathcal{B}$ such that $\mu(\{E,X,E\})=\mu(E\circ Y)$ is satisfied for all events $E$ in $\mathcal{B}$, and the conditional expectation $Y$ is then denoted by $\mu(X|\mathcal{B})$.[8] In general, $\mu(X|\mathcal{B})$ is not uniquely determined and there may be many different versions; $\mu(X|\mathcal{B})$ becomes unique if $\mu$ is faithful on $\mathcal{B}$ (i.e., $0\leq X\in\mathcal{B}$ and $\mu(X)=0$ holds only for $X=0$). Usually, different normal states yield different conditional expectations. If the same element $Y\in\mathcal{B}$ satisfies the equation $\mu(EXE)=\mu(E\circ Y)$ for all events $E$ in $\mathcal{B}$ and for all normal states $\mu$ with $\mathcal{B}\xrightarrow{\mu}X$ and if there is at least one such state, we denote $Y$ by $\mathbb{E}(X|\mathcal{B})$ and call it an objective conditional expectation of $X$ under the sub-algebra $\mathcal{B}$.

Lemma 2.2: Let $\mathcal{A}$ be a JBW algebra with unit $\mathbb{1}$, and let $\mathcal{B}$ be a JBW sub-algebra with $\mathbb{1}\in\mathcal{B}$.
(i) If, for some $X\in\mathcal{A}$, the family of those normal states $\mu$ that satisfy $\mathcal{B}\xrightarrow{\mu}X$ is faithful on $\mathcal{B}$ (i.e., $0\leq Y\in\mathcal{B}$ with $\mu(Y)=0$ for all these states $\mu$ implies that $Y=0$), then there is at most one version of the objective conditional expectation $\mathbb{E}(X|\mathcal{B})$.
(ii) If, for some $X\in\mathcal{A}$, $\mathbb{E}(X|\mathcal{B})$ exists and is unique, then the family of those normal states $\mu$ that satisfy $\mathcal{B}\xrightarrow{\mu}X$ is faithful on $\mathcal{B}$.
(iii) If $X\in\mathcal{A}$ operator-commutes with each $Y\in\mathcal{B}$ and if $\mathbb{E}(X|\mathcal{B})$ exists, then $X$ must lie in $\mathcal{B}$.
(iv) If the family of those normal states $\mu$ on $\mathcal{A}$ with $\mathcal{B}\xrightarrow{\mu}\mathcal{A}$ is faithful on $\mathcal{B}$ and if $\mathbb{E}(X|\mathcal{B})$ exists for each $X\in\mathcal{A}$, the map $X\to\mathbb{E}(X|\mathcal{B})$ is normal on $\mathcal{A}$.

Proof: (i) We assume that both $Y_1$ and $Y_2$ are versions of $\mathbb{E}(X|\mathcal{B})$ and that $\mathcal{B}\xrightarrow{\mu}X$ holds. Then $\mu(E\circ Y_1)=\mu(E\circ Y_2)$ for all events $E\in\mathcal{B}$, hence $\mu(Z\circ Y_1)=\mu(Z\circ Y_2)$ for all $Z\in\mathcal{B}$ by the spectral theorem. Selecting $Z:=Y_1-Y_2$ yields $\mu((Y_1-Y_2)^2)=0$. Since this holds for all states with $\mathcal{B}\xrightarrow{\mu}X$ and since these states are faithful, we get $Y_1=Y_2$.
(ii) We assume that the states $\mu$ with $\mathcal{B}\xrightarrow{\mu}X$ are not faithful on $\mathcal{B}$. I.e., there is a positive element $Y\in\mathcal{B}$ with $Y\neq 0$ and $\mu(Y^2)=0$ for all these states $\mu$. For each such state $\mu$, the Cauchy-Schwarz inequality then implies that $\mu(E\circ Y)=0$ for all events $E\in\mathcal{B}$. Therefore, adding $Y$ to any version of $\mathbb{E}(X|\mathcal{B})$, yields another different version of $\mathbb{E}(X|\mathcal{B})$.
(iii) Since $X$ operator-commutes with each $Y\in\mathcal{B}$, we have $\mathcal{B}\xrightarrow{\mu}X$ for each state $\mu$ on $\mathcal{A}$. With $Y:=\mathbb{E}(X|\mathcal{B})$, we then get: $\mu(\{E,X,E\})=\mu(E\circ Y)$ for all events $E\in\mathcal{B}$ and all normal states $\mu$ on $\mathcal{A}$. Selecting $E=\mathbb{1}$ yields: $\mu(X)=\mu(Y)$ for all normal states $\mu$. Therefore $X=Y\in\mathcal{B}$.
(iv) Let $X_\alpha$ be an increasing net of positive elements in $\mathcal{A}$ with $\sup X_\alpha=X$. Then the $\mathbb{E}(X_\alpha|\mathcal{B})$ are an increasing net as well with $\mathbb{E}(X_\alpha|\mathcal{B})\leq\mathbb{E}(X|\mathcal{B})$ for each $\alpha$, therefore $\sup\mathbb{E}(X_\alpha|\mathcal{B})\leq\mathbb{E}(X|\mathcal{B})$ and

$\mu(\sup \mathbb{E}(X_\alpha|\mathcal{B}))=\sup\mu(\mathbb{E}(X_\alpha|\mathcal{B}))=\sup\mu(X_\alpha)=\mu(X)=\mu(\mathbb{E}(X|\mathcal{B}))$ for all normal states $\mu$ with $\mathcal{B} \xrightarrow{\mu} \mathcal{A}$. The faithfulness then implies that $\sup \mathbb{E}(X_\alpha|\mathcal{B})=\mathbb{E}(X|\mathcal{B})$. □

Part (iii) of Lemma 2.2 implies that the objective conditional expectation $\mathbb{E}(X|\mathcal{B})$ does not have any meaning in the case of classical probabilities, where only the trivial case $X\in\mathcal{B}$ and $\mathbb{E}(X|\mathcal{B})=X$ occurs.

While the conditional expectations $\mu(X|\mathcal{B})$ and $\mathbb{E}(X|\mathcal{B})$ are defined for each single element $X$ in $\mathcal{A}$, a conditional expectation in operator algebra theory[2,10] is usually assumed to be a positive linear map $\pi$ from $\mathcal{A}$ onto the sub-algebra $\mathcal{B}$ with $\pi(\mathbb{1})=\mathbb{1}$ and $\pi(Y)=Y$ for all $Y\in\mathcal{B}$. Such maps $\pi$ shall be called global conditional expectations in the present paper to distinguish them from the other cases. They have the following important property: $\pi(Y\circ X)=Y\circ\pi(X)$ for all $Y\in\mathcal{B}$, $X\in\mathcal{A}$. If $\mu$ is a faithful normal state with $\mathcal{B}\xrightarrow{\mu}\mathcal{A}$, the map $X\to\mu(X|\mathcal{B})$ is such a global conditional expectation and, under the assumptions of Lemma 2.2 (iv), the map $X\to\mathbb{E}(X|\mathcal{B})$ as well.

## 3. The Lüders - von Neumann measurement

For mutually orthogonal events $E_n$ with $\Sigma_n E_n=\mathbb{1}$ and a further event $F$, the formula $\Sigma_n\mu(F|E_n)\mu(E_n)=\mu(F)$ holds for the classical conditional probabilities $\mu(F|E_n)$ with a probability measure $\mu$ (if $\mu(E)=0$, we lay down the general rule $\mu(F|E)\mu(E):=0$ although $\mu(F|E)$ is not defined in this case), but not for their quantum analogue with a state $\mu$. In the quantum case, therefore, the state $\mu_P$, defined by $\mu_P(F):=\Sigma_n\mu(F|E_n)\mu(E_n)$ for any event $F$, is not identical with the original state $\mu$ itself.

With the standard quantum-mechanical Hilbert space formalism, we get $\mu_P(F)=\mu(\Sigma_n E_n F E_n)$. This is the state after a Lüders - von Neumann measurement of an observable with discrete spectrum (or of a family of compatible observables with discrete joint spectrum) and with the eigen-space projections $E_n$, when the physical system was in the initial state $\mu$ prior to the measurement. Von Neumann[13] originally considered only the case when the $E_n$ are minimal events (atoms), i.e. only an observable with a non-degenerate spectrum (or a family of compatible observables with a non-degenerate joint spectrum). This case is called a complete measurement. Lüders[3] later extended von Neumann's measurement process to the case of an incomplete measurement, i.e., the $E_n$ need no longer be atoms and the spectrum (or joint spectrum) may be degenerate.

The Lüders - von Neumann measurement shall now be extended to observables with non-discrete spectra; for this purpose, we consider an abelian JBW sub-algebra $\mathcal{B}$ of the JBW algebra $\mathcal{A}$ with $\mathbb{1}\in\mathcal{B}$ and try to define $\mu_\mathcal{B}(F)$ as a limit via $\mu_P(F)=\Sigma_n\mu(F|E_n)\mu(E_n)$ with the partitions $P:=\{E_n\}$ lying in $\mathcal{B}$ and becoming finer and finer. Since non-countable partitions shall be included, the index $n$ is replaced by the index $\alpha$ [Non-countable partitions exist on a non-separable Hilbert space, but $\mu(E_\alpha)\neq 0$ then holds only for countably many $E_\alpha$].

A partition $P$ in $\mathcal{B}$ is a family of mutually orthogonal events $E_\alpha\neq 0$ in $\mathcal{B}$ with $\Sigma_\alpha E_\alpha=\mathbb{1}$. A partition $P'=\{E_{\alpha'}\}$ is called finer than the partition $P$ ($P'\geq P$) if each $E_\alpha$ is a sum of some of the $E_{\alpha'}$. With this partial ordering $\geq$, the system $\mathcal{P}$ of all partitions in the abelian sub-algebra $\mathcal{B}$ forms a directed set, i.e., for any two partitions in $\mathcal{B}$ there is a third partition in $\mathcal{B}$ which is finer than each of the two.

Now let $\mu$ be a state on $\mathcal{A}$ and let $F$ be any event in $\mathcal{A}$. For each partition $\{E_\alpha\}=P\in\mathcal{P}$, we define $\mu_P(F)=\Sigma_\alpha\mu(F|E_\alpha)\mu(E_\alpha)$. Now the $\mu_P(F)$ indexed by $P$ form a net (or generalized sequence or Moore-Smith sequence) in the unit interval [0,1]. Due to the compactness of the unit interval, this net has at least one accumulation point. If it has only one accumulation point, it converges and we can define $\mu_\mathcal{B}(F):=\lim \mu_P(F)$. This is the generalization to the non-discrete spectrum that we have been looking for, but it need not exist for all events $F$ in $\mathcal{A}$, since the net may not converge.

For any partition $P\in\mathcal{P}$, we have $\mu_P(F)=\mu(\Sigma_{E\in P}\{E,F,E\})$. The $\Sigma_{E\in P}\{E,F,E\}$, indexed by $P\in\mathcal{P}$, form a net in $\{X\in\mathcal{A}|0\leq X\leq\mathbb{1}\}$. This set is compact with regard to the weak topology generated by the normal linear functionals on $\mathcal{A}$. Therefore, again, it has at least one accumulation point. If it has only one accumulation point, it converges and we can define $\mathbb{M}(F|\mathcal{B}):=\lim_{P\in\mathcal{P}}\Sigma_{E\in P}\{E,F,E\}$; then $\mu_\mathcal{B}(F)$ exists for the normal states $\mu$ on $\mathcal{A}$ and $\mu_\mathcal{B}(F)=\mu(\mathbb{M}(F|\mathcal{B}))$.

The so defined $\mathbb{M}$ provides the generalization of the Lüders - von Neumann measurement process to observables with non-discrete spectra (let then $\mathcal{B}$ be the sub-algebra generated by the measured observable or family of compatible observables). We consider $\mathbb{M}(F|\mathcal{B})$ not only for events $F$ in $\mathcal{A}$, but also for any element $X$ in $\mathcal{A}$ and define $\mathbb{M}(X|\mathcal{B}):=\lim_{P\in\mathcal{P}}\Sigma_{E\in P}\{E,X,E\}$ in the case that this net converges. In contrast to the discrete case, the net need not any more converge for all $X$ and it shall now be studied for which $X$ it does. In the next section, it will be seen that there is a close connection to the question when the objective conditional expectations exist, but first some properties of $\mathbb{M}$ will be studied.

Lemma 3.1: Let $\mathcal{A}$ be a JBW algebra with unit $\mathbb{1}$, let $\mathcal{B}$ be an abelian JBW sub-algebra with $\mathbb{1}\in\mathcal{B}$ and let $\mathcal{B}'$ be the commutant of $\mathcal{B}$, i.e., $\mathcal{B}'$ contains all those elements of $\mathcal{A}$ that operator-commute with each element in $\mathcal{B}$.
(i) The net $\Sigma_{E\in P}\{E,X,E\}$, indexed by $P\in\mathcal{P}$, has at least one accumulation point, and each accumulation point $Y$ lies in the commutant $\mathcal{B}'$ and satisfies $\|Y\|\leq\|X\|$.
(ii) If $\mathbb{M}(X|\mathcal{B})$ exists, then $\mathbb{M}(X|\mathcal{B})\in\mathcal{B}'$ and $\|\mathbb{M}(X|\mathcal{B})\|\leq\|X\|$.
(iii) If $\mathbb{M}(X|\mathcal{B})$ exists and $Y\in\mathcal{B}'$, then $\mathbb{M}(Y\circ X|\mathcal{B})$ exists and $\mathbb{M}(Y\circ X|\mathcal{B})=Y\circ\mathbb{M}(X|\mathcal{B})$.
(iv) If $X\in\mathcal{B}'$, then $\mathbb{M}(X|\mathcal{B})$ exists and $\mathbb{M}(X|\mathcal{B})=X$.
(v) If $\mathbb{M}(X|\mathcal{B})$ exists, then $\mathbb{M}(\mathbb{M}(X|\mathcal{B})|\mathcal{B})$ exists and $\mathbb{M}(\mathbb{M}(X|\mathcal{B})|\mathcal{B})=\mathbb{M}(X|\mathcal{B})$.
(vi) Assume that $\mathcal{A}$ is the self-adjoint part of a $W^*$-algebra and that $U$ is any unitary element in this $W^*$-algebra such that $U$ commutes with $\mathcal{B}$. If $\mathbb{M}(X|\mathcal{B})$ exists, then $\mathbb{M}(UXU^{-1}|\mathcal{B})$ exists and $\mathbb{M}(UXU^{-1}|\mathcal{B})=\mathbb{M}(X|\mathcal{B})$.

Proof: (i) Since $\|\Sigma_{E\in P}\{E,X,E\}\|\leq\|X\|$ for each partition $P$ and since the set $\{Z\in\mathcal{A}: \|Z\|\leq\|X\|\}$ is weakly compact, the net $\Sigma_{E\in P}\{E,X,E\}$, indexed by $P\in\mathcal{P}$, has at least one accumulation point in this compact set. We assume that $Y$ is such an accumulation point. Then there is a subnet $\mathcal{Q}$ of $\mathcal{P}$ such that $Y=\lim_{P\in\mathcal{Q}}\Sigma_{E\in P}\{E,X,E\}$.

For any event $D$ in $\mathcal{B}$, there is a partition $P_o$ in $\mathcal{Q}$ which is finer than the partition formed by $D$ and $D'$. Therefore, for each partition $P$ finer than $P_o$, either $E\leq D$ or $E\perp D$ holds for $E\in\mathcal{P}$, and in both cases $D$ and $\{E,F,E\}$ operator-commute. Thus $D$ and $Y$ operator-commute. Since this holds for all events $D$ in $\mathcal{B}$, we get that $Y\in\mathcal{B}'$.

Part (ii) immediately follows from (i). Parts (iii), (iv), and (vi) are direct consequences of the definition of $\mathbb{M}$, and (v) follows from (ii) and (iv). □

Lemma 3.1 (iv) means that only the trivial case $\mathbb{M}(X|\mathcal{B})=X$ occurs in abelian (i.e., classical) algebras, and (v) means that a repetition of the same measurement reproduces the result of the first measurement.

## 4. The existence theorems

The following theorem provides a criterion to decide for which elements $X$ in $\mathcal{A}$ the generalization of the Lüders - von Neumann measurement process and the objective conditional expectation exist.

**Theorem 4.1:** Let $\mathcal{A}$ be a JBW algebra with unit $\mathbb{1}$, let $\mathcal{B}$ be an abelian JBW sub-algebra with $\mathbb{1} \in \mathcal{B}$ and let $\mathcal{B}'$ be the commutant of $\mathcal{B}$. Then the following two conditions are equivalent for an element $X$ in $\mathcal{A}$:

(i) The family of those normal states $\mu$ on $\mathcal{A}$ with $\mathcal{B}' \xrightarrow{\mu} X$ is faithful on $\mathcal{B}'$.

(ii) $\mathbb{E}(X|\mathcal{B}')$ exists and is unique.

If one of these two conditions (i) or (ii) is satisfied, $\mathbb{M}(X|\mathcal{B})$ exists and the identity $\mathbb{M}(X|\mathcal{B}) = \mathbb{E}(X|\mathcal{B}')$ holds.

Proof: By Lemma 3.1, the net $\Sigma_{E \in P}\{E,X,E\}$, indexed by $P \in \mathcal{P}$, has at least one accumulation point $Y$, and $Y \in \mathcal{B}'$. Let $\mu$ be any normal state satisfying $\mathcal{B}' \xrightarrow{\mu} X$. For any partition $P \in \mathcal{P}$ and any event $D$ in $\mathcal{B}'$ we then have:

$$\mu(D \circ \Sigma_{E \in P}\{E,X,E\}) = \mu((\Sigma_{E \in P}D \circ E) \circ (\Sigma_{E \in P}\{E,X,E\})) = \mu(\Sigma_{E \in P}((D \circ E) \circ \{E,X,E\}))$$
$$= \mu(\Sigma_{E \in P}(\{E,X,D \circ E\})) = \mu(\Sigma_{E \in P}(\{D \circ E,X,D \circ E\} + \{D' \circ E,X,D \circ E\}))$$
$$= \Sigma_{E \in P}\mu(\{D \circ E,X,D \circ E\}) = \Sigma_{E \in P}\mu(X \circ (D \circ E)) = \mu(X \circ D) = \mu(\{D,X,D\}).$$

Note that the identity $F \circ \{E,X,E\} = \{E,X,F\}$ holding for events $E$ and $F$ with $F \leq E$ has been used here with $F = D \circ E$ and that Lemma 2.1 (iv) has been applied.

Therefore, $\mu(D \circ Y) = \mu(\{D,X,D\})$ for any event $D$ in $\mathcal{B}'$; i.e., $Y$ is a version of the conditional expectation $\mu(X|\mathcal{B}')$ and, since $Y$ does not depend on $\mu$, $Y$ is a version of $\mathbb{E}(X|\mathcal{B}')$. This holds for any accumulation point $Y$.

We first assume (i). By Lemma 2.2 (i), the faithfulness implies the uniqueness of $\mathbb{E}(X|\mathcal{B}')$ and the coincidence of each accumulation point $Y$ with $\mathbb{E}(X|\mathcal{B}')$. We thus get (ii) as well as the existence of $\mathbb{M}(X|\mathcal{B})$ with the identity $\mathbb{M}(X|\mathcal{B}) = \mathbb{E}(X|\mathcal{B}')$. Condition (ii) implies (i) by Lemma 2.2 (ii). □

While Theorem 4.1 deals with the conditional expectations of single operators, the following theorem considers the global conditional expectations.

**Theorem 4.2:** Let $\mathcal{A}$ be a JBW algebra with unit $\mathbb{1}$, let $\mathcal{B}$ be an abelian JBW sub-algebra with $\mathbb{1} \in \mathcal{B}$ and let $\mathcal{B}'$ be the commutant of $\mathcal{B}$. Then the following conditions (i) and (ii) are equivalent; if, moreover, the commutant $\mathcal{B}'$ is abelian, then all the following four conditions are equivalent:

(i) There is a normal global conditional expectation $\pi: \mathcal{A} \to \mathcal{B}'$ with $\pi(Y) = Y$ for $Y \in \mathcal{B}'$.

(ii) $\mathbb{M}(X|\mathcal{B})$ exists for all $X \in \mathcal{A}$ and the map $X \to \mathbb{M}(X|\mathcal{B})$ is normal on $\mathcal{A}$.

(iii) The family of those normal states $\mu$ on $\mathcal{A}$ with $\mathcal{B}' \xrightarrow{\mu} \mathcal{A}$ is faithful on $\mathcal{B}'$.

(iv) $\mathbb{E}(X|\mathcal{B}')$ exists and is unique for all $X \in \mathcal{A}$. The map $X \to \mathbb{E}(X|\mathcal{B}')$ is normal and linear on $\mathcal{A}$.

If (i) or (ii) is satisfied, there is only one unique global conditional expectation $\pi$ and the identity $\pi(X) = \mathbb{M}(X|\mathcal{B})$ holds for all $X \in \mathcal{A}$. If $\mathcal{B}'$ is abelian and one of these four conditions is satisfied, the identity $\mathbb{E}(X|\mathcal{B}') = \pi(X) = \mathbb{M}(X|\mathcal{B})$ holds for all $X \in \mathcal{A}$.

Proof: We assume (i) and that, for some $X \in \mathcal{A}$, $Y$ is an accumulation point of the net $\Sigma_{E \in P}\{E,X,E\}$, indexed by $P \in \mathcal{P}$, with $\mathcal{Q}$ being a subnet of $\mathcal{P}$ such that $Y=\lim_{P \in \mathcal{Q}} \Sigma_{E \in P}\{E,X,E\}$. Then $Y \in \mathcal{B}'$, $\pi(\Sigma_{E \in P}\{E,X,E\}) = \Sigma_{E \in P}\pi(\{E,X,E\}) = \Sigma_{E \in P}\{E,\pi(X),E\} = \Sigma_{E \in P} E \circ \pi(X) = \pi(X)$ for each partition $P \in \mathcal{P}$, and thus $Y = \pi(Y) = \lim_{P \in \mathcal{Q}} \pi(\Sigma_{E \in P}\{E,X,E\}) = \pi(X)$. Since each accumulation point $Y$ coincides with $\pi(X)$, $\mathbb{M}(X|\mathcal{B})$ exists and the identity $\pi(X) = \mathbb{M}(X|\mathcal{B})$ holds. Condition (i) immediately follows from (ii) by defining $\pi$ as $\pi(X) := \mathbb{M}(X|\mathcal{B})$ for $X \in \mathcal{A}$.

We assume (i) and that $\mathcal{B}'$ is abelian. Therefore $\pi(\{F,X,F\}) = \{F,\pi(X),F\} = F \circ \pi(X) = \pi(F \circ X)$ for $X \in \mathcal{A}$ and $F \in \mathcal{B}'$. Now let $\nu$ be any normal state on $\mathcal{A}$; we then define $\mu(X) := \nu(\pi(X))$ for $X \in \mathcal{A}$ and thus get another normal state $\mu$ with $\mu(\{F,X,F\}) = \mu(F \circ X)$ for $X \in \mathcal{A}$ and $F \in \mathcal{B}'$. I.e., $\mathcal{B}' \xrightarrow{\mu} \mathcal{A}$ holds, and $\mu$ coincides with $\nu$ on the sub-algebra $\mathcal{B}'$. Since the set of all normal states on $\mathcal{A}$ is faithful on $\mathcal{B}'$, we get (iii). Condition (iii) implies (iv) by Theorem 4.1 and Lemma 2.2 (iv). Condition (i) immediately follows from (iv) by defining $\pi$ as $\pi(X) := \mathbb{E}(X|\mathcal{B}')$ for $X \in \mathcal{A}$. □

One of the examples considered in the next section will show that, without the commutant $\mathcal{B}'$ being abelian, the first two conditions of Theorem 4.2 do not imply the last two ones.

## 5. Examples and applications

**Objective conditional probabilities for continuous spectra.** Let $Y$ be an observable with real values. If its spectrum is discrete and non-degenerate, $Y = \Sigma \lambda_n F_n$ where the $\lambda_n$ are the eigen-values and the $F_n$ are atoms, and $\mathbb{P}(E|F_n)$ can be interpreted as $\mathbb{P}(E|Y=\lambda_n)$ for some event $E$, i.e., as the state-independent objective conditional probability of $E$ after a measurement of $Y$ has given the result $\lambda_n$. The question is whether this can be extended to a non-degenerate, but continuous spectrum. Can $\mathbb{P}(E|Y=y)$ be defined then in a reasonable way for $y$ belonging to the spectrum?

A similar procedure for the conditional expectations $\mu(E|\mathcal{B})$ in mathematical probability theory suggests to do this by factorizing $\mathbb{E}(E|\mathcal{B})$ or $\mathbb{M}(E|\mathcal{B})$ as $\mathbb{E}(E|\mathcal{B})=f(Y)$ or $\mathbb{M}(E|\mathcal{B})=f(Y)$ with a measurable function $f$ and to define $\mathbb{P}(E|Y=y):=f(y)$ then, where $\mathcal{B}$ denotes the JBW sub-algebra generated by $Y$. The non-degenerate spectrum means that $\mathcal{B}$ is a maximal abelian sub-algebra (i.e., $\mathcal{B}=\mathcal{B}'$); from Theorem 4.2 we know that we need not bother whether to use $\mathbb{E}(E|\mathcal{B})$ or $\mathbb{M}(E|\mathcal{B})$, since either none exists or both exist and coincide. The factorization is possible when $\mathcal{B}=\{f(Y):f$ is a measurable function$\}$; this holds if $Y$ is of countable type (i.e., each partition that consists of non-zero spectral projections of $X$ is countable), which is always the case in the quantum mechanical standard model with a separable Hilbert space. Thus, it is possible to define $\mathbb{P}(E|Y=y)$ for a non-degenerate continuous spectrum in certain cases, while $\mathbb{P}(E|Y=\lambda_n)$ always exits in the case of a non-degenerate discrete spectrum.

**Algebras with faithful traces.** A trace on a usual operator algebra has the well-known properties $tr(X) \geq 0$ for $X \geq 0$, $tr(X)=tr(UXU^{-1})$ for the unitary operators $U$ and $tr(XY)=tr(YX)$ for all $X,Y$. A trace on a JBW algebra $\mathcal{A}$ has the properties $tr(X) \geq 0$ for $X \geq 0$, $tr(X)=tr(SXS)$ for the symmetries $S$ in $\mathcal{A}$, $tr(X \circ (Y \circ Z))=tr((X \circ Y) \circ Z)$, $tr(X \circ \{Z,Y,Z\})=tr(\{Z,X,Z\} \circ Y)$, and $tr(X \circ Y) \geq 0$ for $X \geq 0$, $Y \geq 0$ ($X,Y,Z \in \mathcal{A}$)

Let $\mathcal{A}$ be a JBW algebra with a faithful normal trace. Let $\mathcal{B}$ be an abelian JBW sub-algebra containing a partition $E_\alpha$ such that each $E_\alpha$ has a finite trace. Then, $\mu_\alpha(X):=tr(E_\alpha \circ X)/tr(E_\alpha)$ for $X \in \mathcal{A}$ defines a faithful family of normal states satisfying $\mathcal{B}' \xrightarrow{\mu} \mathcal{A}$ for $\mu=\mu_\alpha$ and each $\alpha$. By Theorem 4.1, $\mathbb{M}(X|\mathcal{B})$ and $\mathbb{E}(X|\mathcal{B}')$ then exist for all $X \in \mathcal{A}$. By Lemma 2.2, the map $X \to \mathbb{M}(X|\mathcal{B})$ $= \mathbb{E}(X|\mathcal{B}')$ is normal on $\mathcal{A}$.

This situation covers the type I and type II JBW factors; in a type III factor, finite trace partitions do not exist. If the trace on $\mathcal{A}$ is finite (e.g., $\mathcal{A}$ is a type $I_n$ or $II_1$ factor), the trivial partition $\{\mathbb{I}\}$ is a finite trace partition such that $\mathbb{M}(X|\mathcal{B})$ and $\mathbb{E}(X|\mathcal{B}')$ exist for all $X \in \mathcal{A}$ and for any abelian JBW sub-algebra $\mathcal{B}$. In the type $II_1$ case, $\mathbb{P}(E|Y=y)$ exists for all events $E$ and for each observable $Y$ that has a non-degenerate continuous spectrum and is of countable type. This case and the other type II cases where $\mathcal{B}$ does not contain events that are minimal in $\mathcal{B}$ are not covered by the usual Lüders - von Neumann measurement.

**Type I factors.** The standard model of quantum mechanics uses the algebra of all self-adjoint bounded linear operators on the separable Hilbert space with infinite dimension and includes observables with non-degenerate continuous spectra. An important question is whether the usual Lüders - von Neumann measurement can be extended to such observables. However, Theorem 4.2 and part (ii) of the following lemma show that no normal global conditional expectation onto the sub-algebra generated by such an observable exists; note that a non-degenerate spectrum means that this sub-algebra is maximal abelian. Such a no-go result has originally been proved by Areveson[1] for the standard model of quantum mechanics. In the framework of the present paper, we get it for all type I JBW factors.

Lemma 5.1: Let $\mathcal{A}$ be type I JBW factor (or any direct sum of type I JBW factors) and let $\mathcal{B}$ be a sub-algebra with $\mathbb{I} \in \mathcal{B}$.
(i) Let $tr$ denote the standard trace function on $\mathcal{A}$ (sum of the standard trace functions on each factor summand). The state $\mu$ with the shape $\mu(X) = tr(\rho \circ X)$ for $X \in \mathcal{A}$ with some positive $\rho \in \mathcal{A}$, $tr(\rho) = 1$, satisfies $\mathcal{B} \xrightarrow{\mu} \mathcal{A}$ if and only if $\rho \in \mathcal{B}'$.
(ii) Let $\mathcal{B}$ be a maximal abelian sub-algebra (i.e., $\mathcal{B} = \mathcal{B}'$). Then the family of those normal states $\mu$ on $\mathcal{A}$ with $\mathcal{B} \xrightarrow{\mu} \mathcal{A}$ is faithful on $\mathcal{B}$ if and only if $\mathcal{B}$ is generated by atoms (minimal in $\mathcal{A}$).

Proof: (i) If $\rho \in \mathcal{B}'$, then $tr(\rho \circ \{E,X,E\}) = tr(\{E,\rho,E\} \circ X) = tr((E \circ \rho) \circ X) = tr(E \circ (\rho \circ X))$ for $X \in \mathcal{A}$ and events $E \in \mathcal{B}$. We now assume that $\mu(X) = tr(\rho \circ X)$ for $X \in \mathcal{A}$ with some positive $\rho \in \mathcal{A}$, $tr(\rho) = 1$, and that $\mathcal{B} \xrightarrow{\mu} \mathcal{A}$ holds. Since $tr(\{S,\rho,S\} \circ X) = tr(\rho \circ \{S,X,S\}) = tr(\rho \circ X)$ holds for all $X \in \mathcal{A}$ by Lemma 2.1, it follows that $\{S,\rho,S\} = \rho$ for each symmetry $S$ in $\mathcal{B}$. Hence $\rho \in \mathcal{B}'$.

(ii) We first assume that the family of those normal states $\mu$ on $\mathcal{A}$ with $\mathcal{B} \xrightarrow{\mu} \mathcal{A}$ is faithful on $\mathcal{B}$. Let $E \neq 0$ be any event in $\mathcal{B}$. There is a state $\nu$ on $\mathcal{A}$ with $\mathcal{B} \xrightarrow{\nu} \mathcal{A}$ and $\nu(E) > 0$, and we define $\mu(X) := \nu(\{E,X,E\})/\nu(E)$ for $X \in \mathcal{A}$. Then $\mu(E) = 1$ and $\mathcal{B} \xrightarrow{\mu} \mathcal{A}$ holds since $\mathcal{B}$ is abelian and contains $E$ (e.g., use Lemma 2.1 (iv)). The state $\mu$ has the shape $\mu(X) = tr(\rho \circ X)$ for $X \in \mathcal{A}$ with some positive $\rho \in \mathcal{A}$, and from (i) we get that $\rho \in \mathcal{B}' = \mathcal{B}$. Then $\rho$ has at least one non-zero eigenvalue; the spectral projection $D_1$ of $\rho$ belonging to this eigenvalue also lies in $\mathcal{B}$ and has a finite trace because $tr(\rho) = 1$. Thus $tr(D_1) \in \{1, 2, ...\}$. Either $D_1$ is already a minimal projection in $\mathcal{B}$, or there is $D_2$ in $\mathcal{B}$ with $D_2 \leq D_1$ and $tr(D_2) < tr(D_1)$. If $D_2$ is not minimal in $\mathcal{B}$, this process can be continued, but after $n = tr(D)$ steps at latest, a minimal projection $D$ in $\mathcal{B}$ is found. Then $D$ is also minimal in $\mathcal{A}$, since any event below $D$ commutes with $\mathcal{B}$ and thus belongs to $\mathcal{B}$. Since $1 = \mu(E) = tr(\rho E)$, we have $D \leq E$.

Let now $D_\alpha$ be a maximal family of orthogonal atoms in $\mathcal{B}$ (which exists by Zorn's Lemma). If $\mathbb{1}-\Sigma D_\alpha \neq 0$, there is an atom below $\mathbb{1}-\Sigma D_\alpha$, contradicting the maximality of the family $D_\alpha$. Therefore $\mathbb{1}=\Sigma D_\alpha$, and the $D_\alpha$ generate $\mathcal{B}$.

We now assume that $\mathcal{B}$ is generated by atoms, and for each atom $E$ we define a state $\mu_E$ via $\mu_E(X):=tr(E\circ X)$ for $X\in\mathcal{A}$. The family of these states is then faithful on $\mathcal{B}$ and each $\mu=\mu_E$ satisfies $\mathcal{B}\xrightarrow{\mu}\mathcal{A}$.      □

**The canonical commutator relation.** The standard model of quantum mechanics includes observables $X$ and $Y$ both having an unbounded non-degenerate continuous spectrum and satisfying the commutator relation $[X,Y]=i$. Let $\mathcal{B}$ be the weakly closed sub-algebra generated by $X$. Note that then $\mathcal{B}=\mathcal{B}'$. It follows from the result above that neither a normal global conditional expectation $\pi$ onto $\mathcal{B}$ nor $\mathbb{M}(Z|\mathcal{B})$ nor $\mathbb{E}(Z|\mathcal{B})$ exist for <u>all</u> $Z\in\mathcal{A}$. We shall now study the question whether normal states $\mu$ exist that satisfy $\mathcal{B}\xrightarrow{\mu}E$ and whether $\mathbb{M}(E|\mathcal{B})$ or $\mathbb{E}(E|\mathcal{B})$ exist if the event $E$ is one of the spectral projections of $Y$.

For each Borel set $B$ in the real numbers let $E_B$ denote the corresponding spectral projection of $Y$. Moreover define $U_s:=exp(isX)$; $U_s$ is a unitary element in $\mathcal{B}$ for each real number $s$ with $U_s E_B U_s^{-1}=E_{B-s}$. If now a normal state $\mu$ satisfies $\mathcal{B}\xrightarrow{\mu}E_B$, then Lemma 2.1 implies that $\mu(E_B)=\mu(U_s E_B U_s^{-1})=\mu(E_{B-s})$ holds for all real numbers $s$ and all Borel sets $B$. This means that the distribution of $Y$ in the state $\mu$ is invariant under translations and therefore becomes a multiple of the Lebesgue measure. The Lebesgue measure cannot be normalized, which contradicts the fact that the distribution of $Y$ in the state $\mu$ is a probability distribution.

Particularly for the spectral projection $E_I$ belonging to the unit interval $I=]0,1]$ we get $n\mu(E_I)=\Sigma_{k=1,...,n}\mu(E_{I-k})=\mu(E_{]-n,0]})\leq 1$ for all $n$, hence $\mu(E_I)=0=\mu(E_{I-k})$ for all $k$ and finally the contradiction $1=\Sigma_{-\infty<k<\infty}\mu(E_{I-k})=0$. Therefore, no normal state $\mu$ exists satisfying $\mathcal{B}\xrightarrow{\mu}E_I$. The objective conditional expectation $\mathbb{E}(E_I|\mathcal{B})$ is not defined then.

We shall now show that $\mathbb{M}(E_I|\mathcal{B})$ does not exist either. If $\mathbb{M}(E_I|\mathcal{B})$ exists, then $\mathbb{M}(U_s E_I U_s^{-1}|\mathcal{B})$ exists as well and, by Lemma 3.1, $\mathbb{M}(E_I|\mathcal{B})=\mathbb{M}(U_s E_I U_s^{-1}|\mathcal{B})=\mathbb{M}(E_{I-s}|\mathcal{B})$ for any real number $s$. Furthermore, $n\mathbb{M}(E_I|\mathcal{B})=\Sigma_{k=1,...,n}\mathbb{M}(E_{I-k}|\mathcal{B})=\mathbb{M}(E_{]-n,0]}|\mathcal{B})\leq\mathbb{1}$ for all $n$. Since $0\leq\mathbb{M}(E_I|\mathcal{B})$, this implies that $\mathbb{M}(E_I|\mathcal{B})=0$, and we finally get the contradiction $\mathbb{1}=\mathbb{M}(\mathbb{1}|\mathcal{B})=\Sigma_{-\infty<k<\infty}\mathbb{M}(E_{I-k}|\mathcal{B})=\Sigma_{-\infty<k<\infty}\mathbb{M}(E_I|\mathcal{B})=0$, assuming $\sigma$-additivity.

The standard model of quantum mechanics and the observables $X$ and $Y$ as above can also be used to construct an example where $\mathbb{M}(Z|\mathcal{B})$ exists for all $Z\in\mathcal{A}$ and is a normal global conditional expectation onto $\mathcal{B}'$, while there is no normal state $\mu$ with $\mathcal{B}'\xrightarrow{\mu}\mathcal{A}$. We use the same spectral projection $E_I$ of $Y$ as above and define $\mathcal{B}:=\mathbb{R}E_I\oplus\mathbb{R}E_I'$. Then $\mathcal{B}'=E_I\mathcal{A}E_I\oplus E_I'\mathcal{A}E_I'$ and, by von Neumann's bicommutant theorem, $\mathcal{B}''=\mathcal{B}$. Moreover, $\mathbb{M}(Z|\mathcal{B})=E_I Z E_I+E_I' Z E_I'$ for all $Z\in\mathcal{A}$, which is a normal global conditional expectation. However, since $tr(E_I)=\infty=tr(E_I')$, $\mathcal{B}''=\mathcal{B}$ does not contain any $\rho$ with $tr(\rho)=1$ and, by Lemma 5.1 (i), there is no normal state $\mu$ with $\mathcal{B}'\xrightarrow{\mu}\mathcal{A}$.

**Tensor products.** Since a reasonable tensor product for general JBW algebras is not defined, we consider JBW algebras that are the self-adjoint part of W*-algebras. Let $\mathcal{M}$ and $\mathcal{N}$ be W*-algebras, $\mathcal{A}$ the self-adjoint part of $\mathcal{M}\otimes\mathcal{N}$, $\mathcal{B}$ the natural embedding of the self-adjoint part of $\mathcal{M}$ in $\mathcal{A}$ and $X$ the natural embedding of some self-adjoint element of $\mathcal{N}$ in $\mathcal{A}$. Then $X$ operator-commutes with $\mathcal{B}$ such that $\mathbb{E}(X|\mathcal{B})$ does not exist by Lemma 2.2 (iii) [unless $X=\lambda\mathbb{1}$ for some real number $\lambda$].

The conditional expectation μ($X|\mathcal{B}$) exists for each normal state μ on $\mathcal{A}$. If μ is a faithful normal trace state, μ($Y|\mathcal{B}$) exists for each $Y \in \mathcal{A}$, π($Y$):=μ($Y|\mathcal{B}$) defines a normal global conditional expectation, and many other such global conditional expectations ν($Y|\mathcal{B}$) are obtained by using the states ν($Y$):=μ($Z \circ Y$) with $Z \in \mathcal{B}'$. Note that $\mathcal{B}'$ contains the self-adjoint part of $\mathcal{N}$. If $\mathcal{M}$ is abelian, $\mathbb{M}(Y|\mathcal{B})$ exists and $\mathbb{M}(Y|\mathcal{B})=Y$ for all $Y \in \mathcal{A}$, thus coinciding with $\mathbb{E}(Y|\mathcal{B}')$ since $\mathcal{B}'=\mathcal{A}$. This means that, without the commutant $\mathcal{B}'$ being abelian, the conditions (i) or (ii) cannot imply the conditions (iii) or (iv) in Theorem 4.2.